\documentclass[aps,prb,reprint,floatfix,superscriptaddress]{revtex4-2}
\usepackage{amsmath,amsfonts,amssymb} %
\usepackage{mathtools} %
\usepackage{wasysym} %
\usepackage{bm} %
\usepackage{graphicx} %
\usepackage[dvipsnames,table]{xcolor} %
\usepackage{tabularx} %
\usepackage[acronym]{glossaries} %
\usepackage{braket} %
\usepackage{multirow} 
\usepackage{physics}
\usepackage{fancyhdr} %
\usepackage{microtype} %
\usepackage{natmove}

\usepackage{multirow}
\usepackage{makecell}

\usepackage[breaklinks, %
colorlinks, linkcolor=Blue, citecolor=Blue, urlcolor=Blue]{hyperref} %
\usepackage{booktabs}
\usepackage[version=3]{mhchem}

\AtBeginDocument{\usepackage{booktabs}} 

\newcommand{\eq}[1]{Eq.~(\ref{#1})} %
\newcommand{\eqs}[1]{Eqs.~(\ref{#1})} %
\newcommand{\fig}[1]{Fig.~\ref{#1}} %
\newcommand{\Fig}[1]{Figure~\ref{#1}} %
\def\be{\begin{equation}} %
\def\ee{\end{equation}} %

\newcommand{\bea}{\begin{eqnarray}}
\newcommand{\eea}{\end{eqnarray}}

\begin{document}

\title{Global Minimization of Electronic Hamiltonian 1-Norm via Linear Programming in the Block Invariant Symmetry Shift (BLISS) Method }
\author{Smik Patel}
\affiliation{Chemical Physics Theory Group, Department of Chemistry, University of Toronto, Toronto, Ontario M5S 3H6, Canada}
\affiliation{Department of Physical and Environmental Sciences, University of Toronto Scarborough, Toronto, Ontario M1C 1A4, Canada}
\author{Aritra Sankar Brahmachari}
\affiliation{Indian Institute of Science Education and Research (IISER), Kolkata, West Bengal 741246, India}
\author{Joshua T. Cantin}
\affiliation{Chemical Physics Theory Group, Department of Chemistry, University of Toronto, Toronto, Ontario M5S 3H6, Canada}
\affiliation{Department of Physical and Environmental Sciences, University of Toronto Scarborough, Toronto, Ontario M1C 1A4, Canada}
\author{Linjun Wang}
\affiliation{Chemical Physics Theory Group, Department of Chemistry, University of Toronto, Toronto, Ontario M5S 3H6, Canada}
\affiliation{Department of Physical and Environmental Sciences, University of Toronto Scarborough, Toronto, Ontario M1C 1A4, Canada}
\author{Artur F. Izmaylov}
\email{artur.izmaylov@utoronto.ca}
\affiliation{Chemical Physics Theory Group, Department of Chemistry, University of Toronto, Toronto, Ontario M5S 3H6, Canada}
\affiliation{Department of Physical and Environmental Sciences, University of Toronto Scarborough, Toronto, Ontario M1C 1A4, Canada}
\date{\today}
\begin{abstract}
    The cost of encoding a system Hamiltonian in a digital quantum computer as a linear combination of unitaries (LCU) grows with the 1-norm of the LCU expansion. The Block Invariant Symmetry Shift (BLISS) technique reduces this 1-norm by modifying the Hamiltonian action on only the \textit{undesired} electron-number subspaces. Previously, BLISS required a computationally expensive nonlinear optimization that was not guaranteed to find the global minimum. Here, we introduce various reformulations of this optimization as a linear programming problem, which guarantees optimality and significantly reduces the computational cost. We apply BLISS to industrially-relevant homogeneous catalysts in active spaces of up to 76 orbitals, finding substantial reductions in both the spectral range of the modified Hamiltonian and the 1-norms of Pauli and fermionic LCUs. Our linear programming techniques for obtaining the BLISS operator enable more efficient Hamiltonian simulation and, by reducing the Hamiltonian's spectral range, offer opportunities for improved LCU groupings to further reduce the 1-norm.
\end{abstract}

\maketitle

\section{\label{intro_sec}Introduction}

Quantum simulations of chemistry and materials science are one of the most promising applications of quantum computing. Of particular interest in chemistry are simulations of strongly correlated systems, which include homogeneous catalysts and systems with stretched high-order bonds. The electronic structures of these systems are challenging to model using known quantum chemistry methods on classical computers. \cite{mcardleQuantumComputationalChemistry2020, reiherElucidatingReactionMechanisms2017}

The target Hamiltonian we wish to simulate is the electronic structure Hamiltonian in the second quantized form \cite{helgakerMolecularElectronicStructureTheory2014}
\begin{equation}
    \hat{H} = \sum_{ij} h_{ij} \hat{F}_j^i + \sum_{ijkl} g_{ijkl} \hat{F}_j^i \hat{F}_l^k,\label{Helec}
\end{equation}
where $\hat{F}_j^i = \sum_\sigma \hat{a}_{i\sigma}^\dagger \hat{a}_{j\sigma}$ are singlet excitation operators,  $\sigma \in \{\alpha,\beta\}$ are $z$-spin projections, and $h_{ij}$, $g_{ijkl}$ are one- and two-electron integral tensors over spatial orbital indices $i,j,k,l$. \footnote{We write the electronic Hamiltonian in terms of excitation operators. In this notation, the two-electron tensor is $g_{ijkl} = \frac{1}{2}\int \int \dd{\vec{r}_1} \dd{\vec{r}_2} \phi_i^\ast(\vec{r}_1)\phi_j(\vec{r}_1)\frac{1}{|\vec{r}_1 - \vec{r}_2|}\phi_k^\ast(\vec{r}_2)\phi_l(\vec{r}_2)$, and the one-electron tensor is $h_{ij} = \int \dd{\vec{r}} \phi_i^\ast(\vec{r}) \left(-\frac{\nabla^2}{2} - \sum_n \frac{Z_n}{|\vec{r} - \vec{R}_n|}\right) \phi_j(\vec{r}) - \sum_k g_{ikkj}$. Here, $\phi_i$ denotes the single-particle basis, $Z_n$ is the charge of the $n$th nucleus, and $\vec{R}_n$ is the position of the $n$th nucleus.} We focus on the problem of finding the ground state of the target Hamiltonian using the fault-tolerant quantum phase estimation (QPE) algorithm \cite{kitaevQuantumMeasurementsAbelian1995, abramsQuantumAlgorithmProviding1999} and with the walk operator obtained from block encoding of the Hamiltonian. \cite{low2019hamiltonian} To build the block encoding, we use a linear combination of unitaries (LCU) decomposition of the Hamiltonian \cite{childsHamiltonianSimulationUsing2012}
\begin{equation}
    \hat{H} = \sum_j \alpha_j \hat{U}_j, \quad \hat{U}_j^\dagger \hat{U}_j = \hat{1}.
\end{equation}
This approach is known to provide optimal linear scaling in time needed for the Hamiltonian simulation within QPE. \cite{low2019hamiltonian,lowHamiltonianSimulationInteraction2019} Many approaches have been developed for generating an LCU decomposition of the electronic Hamiltonian. \cite{zhaoMeasurementReductionVariational2020,izmaylovUnitaryPartitioningApproach2020,loaizaReducingMolecularElectronic2023a,loaizaMajoranaTensorDecomposition2024,leeEvenMoreEfficient2021,berryQubitizationArbitraryBasis2019,vonburgQuantumComputingEnhanced2021,roccaReducingRuntimeFaultTolerant2024a,oumarouAcceleratingQuantumComputations2024,dekaSimultaneouslyOptimizingSymmetry2024}

All LCU-based simulation methods scale \cite{childsHamiltonianSimulationUsing2012} with the 1-norm of the LCU decomposition
\begin{equation}
    \lambda = \sum_j |\alpha_j|.
\end{equation}
It was previously shown  that the 1-norm of any LCU decomposition of $\hat{H}$ is lower bounded by half of the spectral range of $\hat{H}$: $\lambda \geq \Delta E / 2$, where $\Delta E = E_\text{max} - E_\text{min}$, and $E_\text{max(min)}$ is the largest (lowest) eigenvalue of $\hat{H}$. \cite{loaizaReducingMolecularElectronic2023a} This makes the spectral range of the target electronic Hamiltonian a key limiting factor in the efficiency of quantum simulation algorithms applied to chemistry. This relation between the spectral range and simulation cost motivates the development of methods to engineer Hamiltonians with reduced spectral range, whose simulation via QPE will produce the correct eigenstate and energy. In practice, since calculating the spectral range is at least as hard as the ground state problem, one can instead target the 1-norm of a particular LCU decomposition directly, which is an efficiently computable upper bound to the spectral range. However, we note that a reduction in the 1-norm of a Hamiltonian LCU decomposition will not necessarily correspond to a reduction in the spectral range. \footnote{A simple example is to consider the two-qubit Hamiltonians $\hat{H}_1 = \hat{z}_0 + \hat{z}_1 + \hat{z}_0 \hat{z}_1$, and $\hat{H}_2 = 1.25 \hat{z}_0 + 1.25 \hat{z}_1$, where $\hat{z}_i$ is the Pauli-$z$ operator on qubit $i$. $\hat{H}_1$ satisfies $\Delta E/2 = 2$ and $\lambda = 3$, whereas $\hat{H}_2$ satisfies $\Delta E/2 = 2.5$ and $\lambda = 2.5$.}

To reduce the spectral range of the target Hamiltonian, a block invariant symmetry shift (BLISS) operator $\hat{K}$ can be introduced. By definition, $\hat{K}$ satisfies $\hat{K}\ket{\psi} = 0$ for any eigenstate $\ket{\psi}$ of a set of Hamiltonian symmetries $\{\hat{S}_l\}_{l=1}^{L}$, $\hat{S}_l \ket{\psi} = s_l \ket{\psi}$. \cite{loaizaBlockInvariantSymmetryShift2023} The BLISS operator does not modify the eigenvalues of states in the symmetry subspace $\{s_l\}_{l=1}^{L}$, but it modifies the rest of the Hamiltonian spectrum. 
To lower the spectral range of $\hat{H} - \hat{K}$, the BLISS operator $\hat{K}$ is optimized to minimize the 1-norm of the Pauli product LCU decomposition of $\hat{H} - \hat{K}$. Such a heuristic choice of the cost function is motivated by simplicity and its success in practice to reduce the spectral range. A lower bound for the spectral range of $\hat{H} - \hat{K}$ is that of $\hat{H}$ restricted to the symmetry subspace $\{s_l\}_{l=1}^{L}$, \cite{cortesAssessingQueryComplexity2024a} which can be achieved in principle by constructing the operator $\hat{H}_e \hat{P}$, where $\hat{P}$ is the projection operator onto the target symmetric subspace. However, the projection $\hat{P}$ is in general a complicated $N$-electron operator, and therefore is not usable in practice. \cite{yenExactApproximateSymmetry2019}

To find the optimal BLISS operator, a nonlinear gradient-based optimization of the BLISS operator parameters was performed. \cite{loaizaBlockInvariantSymmetryShift2023} Although the resulting shifted Hamiltonians were found to have 1-norm of the Pauli decomposition that is close to the global minimum, such an optimization is not guaranteed to find the global minimum. Furthermore, the computational cost of the optimization is quite steep, rendering this method only applicable to smaller systems. In this work, we introduce methods to find the optimal BLISS operator for minimization of LCU 1-norm using linear programming, which circumvents the cost and local minima problems that arise when using nonlinear optimization. \cite{bertsimas-LPbook} Each technique developed targets a particular LCU decomposition, and guarantees convergence of the 1-norm to the global minimum for the associated LCU decomposition. Note that we have  placed a glossary of method terms in Appendix \ref{appendix_glossary} for the reader.

\section{\label{theory_sec}Obtaining BLISS operators using Linear Programming}

All the linear programs considered in this work for reducing the spectral range and LCU 1-norms of the electronic Hamiltonian target the 1-norm of a particular LCU decomposition and can therefore be cast as the minimization of a cost function of the form $C(\vec{x}) = |A\vec{x} - \vec{b}|_1$, where $\vec{x} \in \mathbb{R}^n, \vec{b} \in \mathbb{R}^m$, $A$ is a linear transformation from $\mathbb{R}^n$ to $\mathbb{R}^m$, and $|\vec{v}|_1 = \sum_i |v_i|$ is the 1-norm. In Appendix \ref{appendix_A}, we show how to present this optimization as a linear program and focus here only on its application  to Pauli and fermionic representations of the Hamiltonian.

Since our focus is the electronic Hamiltonian, which commutes with the number operator, we use the same form of the BLISS operator as in  Ref. \cite{loaizaBlockInvariantSymmetryShift2023} 
\begin{align}
    \hat{K}(\vec{\mu}, \vec{\xi}) &= \mu_1 \left(\hat{N}_e - N_e\right) + \mu_2 \left(\hat{N}_e^2 - N_e^2\right)\nonumber\\
    &+ \sum_{ij} \xi_{ij} \hat{F}^i_j \left(\hat{N}_e - N_e\right)\label{eqn_bliss_op}
\end{align}
where $\hat{N}_e = \sum_{i\sigma} \hat{n}_{i\sigma}$ is the total number operator, and $\hat{n}_{i\sigma} = \hat{a}_{i\sigma}^\dagger \hat{a}_{i\sigma}$ is the number operator on spatial-orbital $i$ with corresponding spin $\sigma$. This operator satisfies $\hat{K}\ket{\psi} = 0$ for any state with $N_e$ electrons. Although the electronic Hamiltonian has other one- and two-electron symmetries, such as $\hat{S}_z, \hat{S}^2, \hat{S}_z^2$, and $\hat{S}_z \hat{N}_e$, incorporation of these symmetries were empirically found to give insignificant improvements in 1-norm and spectral range. Operators of the form $\hat{K}(\vec{\mu}, \vec{0})$, which were originally studied in Ref. \cite{loaizaReducingMolecularElectronic2023a}, are called symmetry shift operators, and these have the additional property that they commute with the Hamiltonian. 

\subsection{\label{theory_pauli_subsec}Pauli Product LCUs}

In the BLISS framework, the BLISS operator $\hat{K}$ is subtracted from the electronic Hamiltonian to reduce its spectral range. The resulting Hamiltonian can be written as 
\begin{equation}
    \hat{H} - \hat{K}(\vec{\mu}, \vec{\xi}) = \sum_{ij} \Tilde{h}_{ij} \hat{F}_j^i + \sum_{ijkl} \Tilde{g}_{ijkl} \hat{F}_j^i \hat{F}_l^k,
\end{equation}
where the modified one- and two-electron integrals are
\begin{align}
     \tilde{h}_{i j}&=h_{i j}-\mu_1 \delta_{i j} +N_e \xi_{i j} \label{lpbliss_1e} \\
     \Tilde{g}_{i j k l}&=g_{i j k l}-\mu_2 \delta_{i j}\delta_{k l}-\frac{1}{2}\big(\xi_{i j}\delta_{k l}+\delta_{i j}\xi_{k l}\big).\label{lpbliss_2e}
\end{align}
To optimally reduce the spectral range of $\hat{H} - \hat{K}$, it was proposed in \cite{loaizaBlockInvariantSymmetryShift2023} to choose the parameters of the BLISS operator $\hat{K}$ to minimize the 1-norm of $\hat{H} - \hat{K}$ when expressed as a linear combination of Pauli products, which can be done via a fermion-to-qubit mapping like the Jordan-Wigner \cite{jordanUeberPaulischeAequivalenzverbot1928} or Bravyi-Kitaev transformations. \cite{bravyiFermionicQuantumComputation2002, seeleyBravyiKitaevTransformationQuantum2012} In both mappings, the 1-norm of the associated Pauli product LCU can be expressed in terms of the modified integrals as follows \cite{koridonOrbitalTransformationsReduce2021}
\begin{align}
        \lambda_\text{Pauli} &= \sum_{i j}\left|\Tilde{h}_{i j}+2 \sum_k \Tilde{g}_{i j k k}\right|\nonumber \\
        &+ \frac{1}{2} \sum_{i j k l}\left|\Tilde{g}_{i j k l}\right| +\sum_{i>k, j>l} \left|\Tilde{g}_{i j k l}-\Tilde{g}_{ilkj}\right|.  \label{eq:norm}
\end{align}
This defines the objective function which we minimize using linear programming. We denote this method, in which the BLISS operator is obtained from minimizing \eq{eq:norm} using linear programming, and then subsequently applied to the electronic Hamiltonian, as linear-programming BLISS (LP-BLISS).

\subsection{\label{theory_fermionic_subsec}Fermionic LCUs}
\label{sec_fermionic_lcu}

In addition to the Pauli product LCU, we can alternatively obtain the BLISS operator by minimizing the 1-norm of an LCU decomposition obtained directly in the fermionic algebra. Before deriving the form of the BLISS operator, we first review the fermionic LCUs relevant to this work and their associated 1-norms. In what follows we use $\hat{H}_{1e}$ and $\hat{H}_{2e}$ to denote the one- and two-electron parts of the electronic Hamiltonian.

\subsubsection{Review of Fermionic LCUs}
Fermionic LCUs are written in terms of reflection operators $\hat{U}^\dagger \hat{r}_{i\sigma} \hat{U}$, where $\hat{U}$ is an orbital rotation and $\hat{r}_{i\sigma} = 1 - 2\hat{n}_{i\sigma}$. \cite{loaizaReducingMolecularElectronic2023a} To obtain a fermionic LCU decomposition of $\hat{H}$, it is convenient to start with a Cartan sub-algebra (CSA) decomposition of the two-electron operator $\hat{H}_{2e}$. \cite{yenCartanSubalgebraApproach2021} This takes the form
\begin{equation}
    \hat{H}_{2e} = \sum_\alpha \hat{H}^{(\alpha)} = \sum_\alpha \hat{U}^{(\alpha)\dagger} \left(\sum_{ij\sigma\tau} \lambda_{ij}^{(\alpha)} \hat{n}_{i\sigma}\hat{n}_{j\tau}\right) \hat{U}^{(\alpha)}, \label{csa_fragments}
\end{equation}
where $\hat{U}^{(\alpha)}$'s are orbital rotations. Conversion of occupation number operators to reflections produces 
\begin{equation}
    \hat{H}^{(\alpha)} = \hat{H}^{(\alpha,\text{LCU})} + \hat{L}_{1e}^{(\alpha)} + c^{(\alpha)},
\end{equation}
where $\hat{H}^{(\alpha,\text{LCU})}$ is the LCU decomposition of $\hat{H}^{(\alpha)}$
\begin{equation}
    \hat{H}^{(\alpha,\text{LCU})} = \hat{U}^{(\alpha)\dagger}\left(\frac{1}{4}\sum_{ij\sigma\tau} \lambda_{ij}^{(\alpha)} \hat{r}_{i\sigma} \hat{r}_{j\tau}\right)\hat{U}^{(\alpha)},
\end{equation}
$\hat{L}_{1e}^{(\alpha)}$ is a one-electron correction operator
\begin{equation}
    \hat{L}_{1e}^{(\alpha)} = \hat{U}^{(\alpha)\dagger} \left(2\sum_{ij\sigma} \lambda_{ij}^{(\alpha)} \hat{n}_{i\sigma}\right) \hat{U}^{(\alpha)},
\end{equation}
and $c^{(\alpha)}$ is an unimportant constant. 
Therefore, the 1-norm of the LCU associated to the fragment $\hat{H}^{(\alpha)}$ can be expressed as
\begin{align}
    \lambda_{\text{CSA}}^{(\alpha)} &= \frac{1}{4}\sum_{i\sigma\not=j\tau} \left|\lambda_{ij}^{(\alpha)}\right|\nonumber\\
    &= \sum_{i\not=j}\left|\lambda_{ij}^{(\alpha)}\right| + \frac{1}{2}\sum_i \left|\lambda_{ii}^{(\alpha)}\right|,\label{fr_1norm}
\end{align}
where we have accounted for the involutory property $\hat{r}_{i\sigma}^2 = \hat{1}$.

Combining the one-electron corrections $\hat{L}_{1e}^{(\alpha)}$ with $\hat{H}_{1e}$ produces a modified one-electron Hamiltonian $\tilde{H}_{1e} = \hat{H}_{1e} + \sum_\alpha \hat{L}_{1e}^{(\alpha)}$. Diagonalizing $\tilde{H}_{1e}$  with an orbital transformation $\tilde{V}$ produces the LCU decomposition
\begin{equation}
    \tilde{H}_{1e} = \tilde{V}^\dagger \left(\frac{1}{2}\sum_{i\sigma} \gamma_i \hat{r}_{i\sigma}\right)\tilde{V} + \tilde{c},
\end{equation}
where $\tilde{c}$ is an unimportant constant, and $\gamma_i$ are the eigenvalues of the one-body-tensor of $\tilde{H}_{1e}$. The corresponding 1-norm of the LCU is 
\begin{equation}
    \lambda_{1e} = \sum_i \left|\gamma_i\right|.\label{onee_1norm}
\end{equation}

A special case of the CSA decomposition is the double factorization (DF) decomposition \cite{pengHighlyEfficientScalable2017, mottaLowRankRepresentations2021, hugginsEfficientNoiseResilient2021} in which the coefficient matrices $\lambda_{ij}^{(\alpha)}$ of the fragments $\hat{H}^{(\alpha)}$ are rank 1, implying that they can be written as an outer product $\lambda_{ij}^{(\alpha)} = \epsilon_i^{(\alpha)} \epsilon_j^{(\alpha)}$ of a vector $\epsilon_i^{(\alpha)}$.  There are two-benefits to assuming a low-rank form of the $\lambda_{ij}^{(\alpha)}$ matrices. First, the decomposition can be obtained via a singular value decomposition (SVD) of the two-body-tensor $g_{ijkl}$ of $\hat{H}_{2e}$, rather than the expensive nonlinear gradient-based optimization necessary to obtain the full-rank fragments. Second, the low-rank form of the $\lambda_{ij}^{(\alpha)}$ allows the fragment to be written as follows
\begin{equation}
    \hat{H}^{(\alpha)} = \hat{U}^{(\alpha)\dagger} \left(\sum_{i\sigma} \epsilon_{i}^{(\alpha)} \hat{n}_{i\sigma}\right)^2 \hat{U}^{(\alpha)}.\label{df_frag}
\end{equation}
The perfect square nature of $\hat{H}^{(\alpha)}$  allows one to use the qubitization procedure of Ref. \cite{vonburgQuantumComputingEnhanced2021} for the block-encoding, which exploits the low-rank structure of the fragments, and for which the associated 1-norm is
\begin{equation}
    \lambda_{\text{DF}}^{(\alpha)} = \frac{1}{2}\left(\sum_i \left|\epsilon_i^{(\alpha)}\right|\right)^2.\label{lr_1norm}
\end{equation}
This quantity is empirically much lower than the 1-norm associated to the full-rank fragments, in which each reflection $\hat{r}_{i\sigma}$, $\hat{r}_{i\sigma} \hat{r}_{j\tau}$ is implemented individually. Therefore, it is beneficial to develop methods which preserve the low-rank property of the DF fragments.

\subsubsection{Applying BLISS to individual fermionic fragments}
\label{section_individual_ferm_frag}

We now introduce two approaches to apply BLISS directly to the fermionic fragments themselves, as was done in the symmetry-compressed double factorization (SCDF) method of Ref. \cite{roccaReducingRuntimeFaultTolerant2024a}. Our approaches differ from SCDF in two ways. First, while SCDF only exploits symmetry shifts (the $\mu_1$ and $\mu_2$ terms of \eq{eqn_bliss_op}), our methods use the full BLISS operator to reduce the 1-norms of fermionic LCUs. Second, our linear programming approach to obtain the BLISS operator is compatible with the SVD method to generate the LCU, whereas the SCDF method uses a nonlinear gradient-based optimization to obtain both the BLISS operator and the LCU. 

Targeting individual fermionic fragments is beneficial for two reasons. First, one can optimize the BLISS operator to lower the 1-norm of the fermionic LCU directly, rather than the 1-norm of the Pauli product LCU. Second, as described above, fermionic LCUs are obtained by first decomposing the Hamiltonian into simple Hermitian fragments and then decomposing the fragments into LCUs. In this case, not only is the spectral range $\Delta E$ of $\hat{H}$ relevant as a lower bound on the 1-norm, but the spectral ranges $\Delta E^{(\alpha)}$ of the individual fragments $\hat{H}^{(\alpha)}$ are also relevant. That is, if $\lambda$ is the 1-norm of $\hat{H}$, and $\lambda^{(\alpha)}$ is the 1-norm of the fragment $\hat{H}^{(\alpha)}$, then the triangle inequality implies
\begin{equation}\label{eq:TEF}
    \frac{\Delta E}{2} \leq \sum_\alpha \frac{\Delta E^{(\alpha)}}{2} \leq \sum_\alpha \lambda^{(\alpha)} = \lambda.
\end{equation}
This shows that we can obtain a more accurate lower bound on $\lambda$ from the spectral ranges of the fragments themselves, rather than from the spectral range of $\hat{H}$. For a low-rank fragment $\hat{H}^{(\alpha)}$, the lower-bound $\Delta E^{(\alpha)} / 2$ coincides with the 1-norm $\lambda_{\text{DF}}^{(\alpha)}$ associated to the qubitization procedure of Ref. \cite{vonburgQuantumComputingEnhanced2021}; for a general full-rank fragment, the 1-norm $\lambda_{\text{CSA}}^{(\alpha)}$ is strictly larger than the lower-bound. Therefore, any deviation of the 1-norm $\lambda$ to $\Delta E/2$ when using low-rank fragments only comes from the requirement to use multiple fragments to generate the LCU. To reduce the lower bound based on the fragment spectral ranges $\Delta E^{(\alpha)}$, one can modify each fragment $\hat{H}^{(\alpha)}$ with a BLISS operator $\hat{K}^{(\alpha)}$. The BLISS operator should be optimized such that the spectral range $\Delta \tilde{E}^{(\alpha)}$ of the modified fragment $\hat{H}^{(\alpha)} - \hat{K}^{(\alpha)}$ is minimized, so that $\Delta \tilde{E}^{(\alpha)} \leq \Delta E^{(\alpha)}$. Then, an LCU decomposition of $\hat{H} - \hat{K}$, where $\hat{K} = \sum_\alpha \hat{K}^{(\alpha)}$, will have a reduced lower bound of $\sum_\alpha \Delta \tilde{E}^{(\alpha)} / 2$. 

To match the form of the Hamiltonian fragments, we also express the BLISS operator in CSA form. For the symmetry shift operators, defined by parameters $\mu_1$ and $\mu_2$, we exploit the fact that orbital rotations commute with the total number operator $\hat{N}_e$. This implies that, for all orbital rotations $\hat{U}$, we have
\begin{align}
    \mu_1 \hat{N}_e &= \hat{U}^\dagger\left(\sum_{i\sigma} \mu_1 \hat{n}_{i\sigma}\right)\hat{U}\label{num_csa}\\
    \mu_2 \hat{N}_e^2 &= \hat{U}^\dagger \left(\sum_{ij\sigma\tau} \mu_2 \hat{n}_{i\sigma}\hat{n}_{j\tau}\right)\hat{U}.\label{numsq_csa}
\end{align}

For the remaining term defined by parameters $\vec{\xi}$, we start with the form $\hat{B}(\vec{\xi}) = \sum_{ij\sigma} \xi_{ij} \hat{a}_{i\sigma}^\dagger \hat{a}_{j\sigma} (\hat{N}_e - N_e)$, which is parametrized by the coefficient matrix $\xi$ of $\hat{O}_{1e} = \sum_{ij\sigma} \xi_{ij} \hat{a}_{i\sigma}^\dagger \hat{a}_{j\sigma}$. Diagonalizing $\hat{O}_{1e}$ by an orbital rotation:
\begin{equation}
    \hat{O}_{1e} = \hat{U}^\dagger \left(\sum_{i\sigma} \theta_{i}\hat{n}_{i\sigma}\right) \hat{U}
\end{equation}
and performing some algebra produces an alternative parametrization
\begin{align}
    \hat{B}(\hat{U}, \vec{\theta}) &= \hat{U}^\dagger\Bigg(\sum_{i\sigma\not=j\tau} \frac{\theta_i + \theta_j}{2}\hat{n}_{i\sigma} \hat{n}_{j\tau}\nonumber\\
    &+ (1 - N_e)\sum_{i\sigma} \theta_i\hat{n}_{i\sigma}\Bigg)\hat{U}\label{bliss_csa}
\end{align}
in terms of spatial-orbital rotations $\hat{U}$ and the vector $\vec{\theta}$ that defines the coefficients of the occupation number operators. 

We note that one can also convert BLISS operators constructed from $\hat{S}_z$ to CSA form using the same procedure, with the additional constraint that the orbital rotations commute with $\hat{S}_z$. This is automatically satisfied for the fragments obtained from low-rank and full-rank decompositions of the electronic Hamiltonian. However, such BLISS operators break the $\alpha \leftrightarrow \beta$ spin symmetry and would require us to index over \textit{spin} orbitals, rather than \textit{spatial} orbitals. Moreover as with the Pauli 1-norm, we did not see an improvement in 1-norms when including $\hat{S}_z$-based BLISS operators; we thus omit the details.

With the CSA form of the BLISS operator, we can apply BLISS to the fermionic fragments to reduce their 1-norm while simultaneously preserving the exact solvability of the modified fragment. With the one-electron operator  $\hat{H}_{1e} = \hat{U}^\dagger \sum_{i\sigma} \epsilon_i \hat{n}_{i\sigma}\hat{U}$, the optimal BLISS modified Hamiltonian takes the form \cite{loaizaReducingMolecularElectronic2023a}
\begin{equation}
    \hat{H}_{1e} - \mu_1 \hat{N}_e = \hat{U}^\dagger \sum_{i\sigma}(\epsilon_i - \mu_1) \hat{n}_{i\sigma} \hat{U}. \label{onee_shift}
\end{equation}
Based on \eq{onee_1norm} for the corresponding 1-norm, we  form a linear program and obtain the optimal solution at $\mu_1^{(\text{opt})} = \text{median}\{\epsilon_i\}$.

For the two-electron fragments $\hat{H}^{(\alpha)}$, the general BLISS operators take the form
\begin{equation}
    \hat{K}^{(\alpha)}(\mu_2^{(\alpha)}, \vec{\theta}^{(\alpha)}) = \mu_2^{(\alpha)} \big(\hat{N}_e^2 - N_e^2\big)+ \hat{B}(\hat{U}^{(\alpha)}, \vec{\theta}^{(\alpha)})\label{fermionic_bliss}.
\end{equation}
Using these BLISS operators, we directly modify each fragment:
\begin{equation}
    \hat{H}^{(\alpha)}(\mu_2^{(\alpha)}, \vec{\theta}^{(\alpha)}) = \hat{H}^{(\alpha)} -  \hat{K}^{(\alpha)}(\mu_2^{(\alpha)}, \vec{\theta}^{(\alpha)}). \label{df_lrbs}
\end{equation}
To obtain values for the parameters $\mu_2^{(\alpha)}, \vec{\theta}^{(\alpha)}$ defining each BLISS operator, we target the 1-norm of the modified fragment based on \eq{fr_1norm} 
\begin{align}
    \lambda_{\text{CSA}}^{(\alpha)}[\mu_2^{(\alpha)}, \vec{\theta}^{(\alpha)}] &= \sum_{i\not=j} \left|\lambda_{ij}^{(\alpha)} - \mu_2^{(\alpha)} - \frac{\theta_i^{(\alpha)} + \theta_j^{(\alpha)}}{2}\right|\nonumber\\
    &+ \frac{1}{2}\sum_i \left|\lambda_{ii}^{(\alpha)} - \mu_2^{(\alpha)} - \theta_i^{(\alpha)}\right|,
\end{align}
which we minimize using linear programming. 

Although the above approach can be used to lower the 1-norm of full-rank or low-rank fragments, it breaks the perfect-square property of the low-rank fragments. Therefore, when this BLISS operator is applied to low-rank fragments, the potential for a large reduction in the 1-norm is mitigated by the inability to use the qubitization approach of Ref. \cite{vonburgQuantumComputingEnhanced2021} for block-encoding of the modified fragment, whose 1-norm would be defined by \eq{lr_1norm}, and the requirement to use an alternative approach whose 1-norm is defined by \eq{fr_1norm}. To avoid this obstacle, we introduce a different modification for low-rank fragments: a single parameter $\phi^{(\alpha)}$ per fragment. This takes the form
\begin{equation}
    \hat{H}^{(\alpha)}(\phi^{(\alpha)}) = \hat{U}^{(\alpha)\dagger} \left(\sum_{i\sigma} (\epsilon_i^{(\alpha)} - \phi^{(\alpha)}) \hat{n}_{i\sigma}\right)^2 \hat{U}^{(\alpha)}. \label{df_lrps}
\end{equation}
As shown in Appendix \ref{appendix_B}, $\hat{H}^{(\alpha)} - \hat{H}^{(\alpha)}(\phi^{(\alpha)})$ is equal to a sum with three contributions: 1) a BLISS operator, 2) a one-electron operator, and 3) a constant.  Therefore, one can generate an LCU decomposition in which the low-rank fragments have a reduced 1-norm by optimizing the parameter $\phi^{(\alpha)}$ for all fragments, as long as one accounts for the resulting one-electron contribution. Based on \eq{lr_1norm}, the 1-norm optimization is a linear program with an analytical solution: $\phi^{(\alpha)} = \text{median}\{\epsilon_{i}^{(\alpha)}\}$. In general there is either a unique median, or infinitely many medians to choose from, but in both cases one can always force $\phi^{(\alpha)}$ to equal a particular $\epsilon_i^{(\alpha)}$, and in doing so remove at least two unitaries $\hat{r}_{i\alpha}, \hat{r}_{i\beta}$ from the one-electron operator defining the LCU decomposition of $\hat{H}^{(\alpha)}(\phi^{(\alpha)})$.  

We have thus defined two methods for 1-norm reduction of fermionic LCUs, both of which use \eq{onee_shift} for the one-electron fragments. For the two-electron fragments, the double-factorization with low-rank \textit{preserving} shifts (DF+LRPS) method is based on  \eq{df_lrps} and the double factorization with low-rank \textit{breaking} shifts (DF+LRBS) method is based on \eq{df_lrbs}. Both of these methods also generate a ``global'' BLISS operator, which is the sum of BLISS operators obtained for each fragment. Therefore, these procedures can also be used as an alternative to LP-BLISS for generating the BLISS operator that is applied directly to the electronic Hamiltonian. That is, we again produce modified one- and two-body integrals of the form shown in \eqs{lpbliss_1e} and (\ref{lpbliss_2e}). We call these methods for generating the BLISS operator fermionic-low-rank BLISS (FLR-BLISS) and fermionic-full-rank BLISS (FFR-BLISS) respectively.

\section{\label{results_sec}Results and Discussion}

Here we assess the capability of linear programming techniques for reducing the spectral ranges and LCU 1-norms of electronic Hamiltonians. We focus here on the set of transition-metal homogeneous nitrogen fixation catalysts studied in Ref. \cite{bellonziFeasibilityAcceleratingHomogeneous2024}, as well as the two active spaces of the FeMo cofactor of the nitrogenase enzyme (FeMoco) introduced in Refs. \cite{reiherElucidatingReactionMechanisms2017} and \cite{liElectronicComplexityGroundstate2019}. We choose the nitrogen fixation catalysts as they have been shown to be relevant to industrial applications. \cite{bellonziFeasibilityAcceleratingHomogeneous2024} Furthermore, these molecules, along with FeMoco, are difficult to simulate on classical computers and cover a variety of system sizes and number of electrons; see Table \ref{table_catalysts} for details. To solve the linear program necessary to obtain the LP-BLISS operator, we use the Julia package JuMP.jl. \cite{Lubin2023} To solve the linear programs in the DF+LRBS method and to obtain the FFR-BLISS operator, we used the CVXOPT package \cite{CVXOPTCite} in Python with the GLPK solver. \cite{GLPKCite} Note that we have placed a glossary of BLISS method terms in Appendix \ref{appendix_glossary}, and that all raw numerical data can be found in the Supplementary Information.

\begin{table}
\centering
{\begin{tabularx}{\columnwidth}{@{\extracolsep{\fill}} l l c c }
\toprule
Catalyst System & Molecule ID & $N_\text{orb}$ & $N_e$ \\ \midrule
\multirow{2}[-4]{3.5cm}{FeMoco} & FeMoco Sm  & 54 & 54 \\ 
                                & FeMoco Lg  & 76 & 113 \\ \midrule
\multirow{4}[-11]{3.5cm}{Schrock Catalyst}   & MoN$_2$      & 30 & 45 \\ 
                                            & MoN$_2^-$        & 31 & 46 \\ 
                                            & Fe(Cp)$_2$   & 46 & 58 \\ 
                                            & Fe(Cp)$_2^+$     & 46 & 57 \\ \midrule
\multirow{3}[-6]{3cm}{Bridged Dimolybdenum} & 1-Lut$_{Re}$   & 69 & 90 \\ 
                                            & 1-Lut$_{TS}$   & 69 & 90 \\ 
                                            & II-Lut$_{Prod}$ & 70 & 90 \\ \midrule
\multirow{8}[-21]{3.5cm}{Molybdenum Pincer (small active space)}           & RC Sm & 31     & 44 \\ 
                                                                          & TS$_{1/2}$ Sm         & 31 & 44 \\ 
                                                                          & PC Sm & 31     & 44 \\ 
                                                                          & 2 Sm & 31      & 44 \\ 
                                                                          & I Sm & 55     & 73 \\ 
                                                                          & TS$_{\text{I/4a}}$ Sm & 55 & 73 \\ 
                                                                          & PC$^-$ Sm & 55 & 73 \\ 
                                                                          & 4a Sm & 25     & 37 \\ \midrule
\multirow{8}[-21]{3.5cm}{Molybdenum Pincer (large active space)}          & RC Lg & 45     & 64 \\ 
                                                                          & TS$_{1/2}$ Lg         & 45 & 64 \\ 
                                                                          & PC Lg& 45      & 64 \\ 
                                                                          & 2 Lg &45      & 64 \\ 
                                                                          & I Lg& 69       & 93 \\ 
                                                                          & TS$_{\text{I/4a}}$ Lg & 70 & 93 \\ 
                                                                          & PC$^-$ Lg & 69 & 93 \\ 
                                                                          & 4a Lg & 39    & 57 \\ \bottomrule
\end{tabularx}}
\caption{Active space and molecule properties of the nitrogen fixation catalyst Hamiltonians considered in this work. ``Sm'' and ``Lg'' refer to the smaller and larger active space versions of a given system. $N_\text{orb}$ denotes the number of orbitals.}
\label{table_catalysts}
\end{table}

\begin{figure*}
 	\centering
    \hspace{-0.8cm}
 	\includegraphics[width=0.85\textwidth]{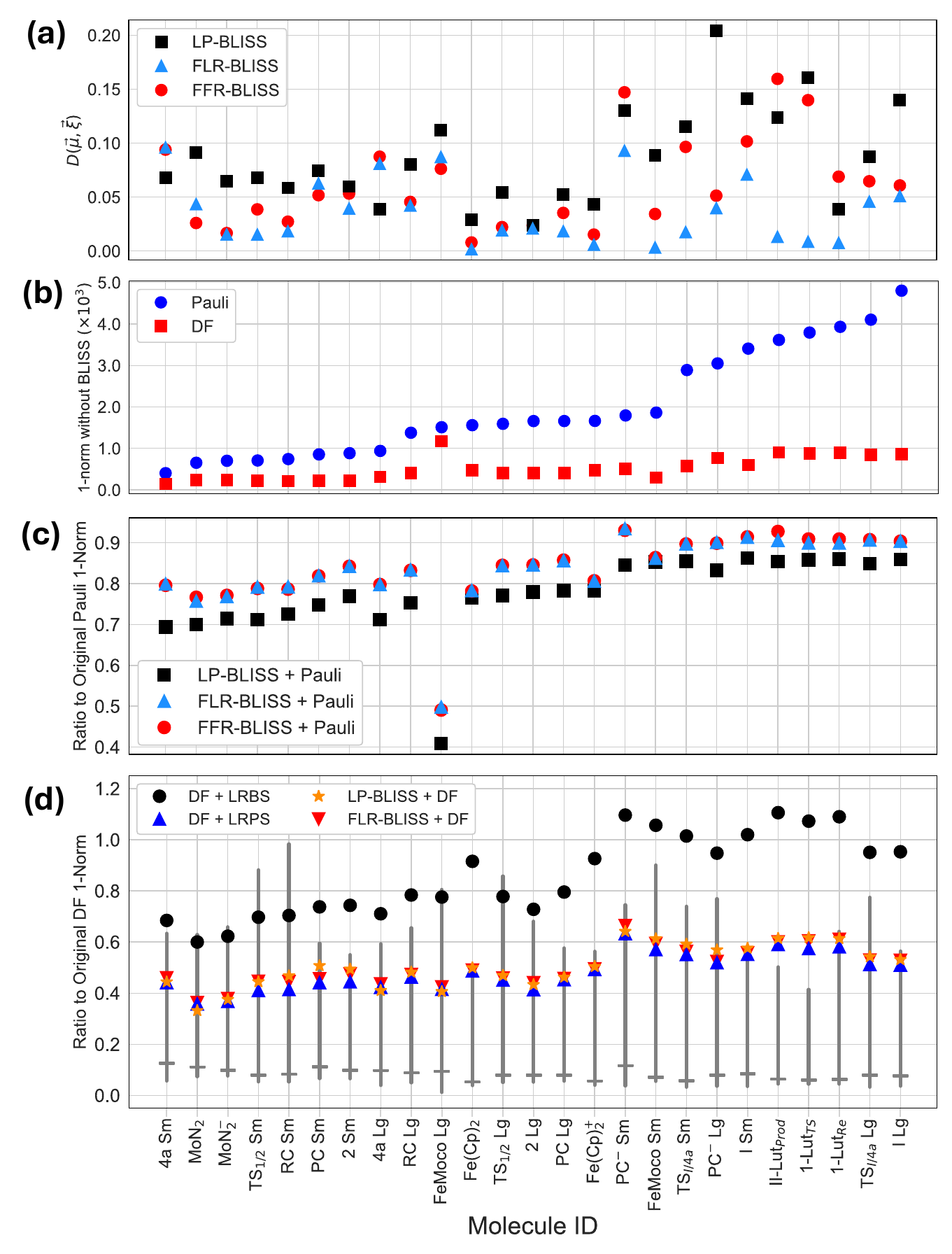}
 	\caption{Spectral range and 1-norm data with and without BLISS operators. The molecules are sorted by their Pauli LCU 1-norms. (a) Normalized deviation $D(\vec{\mu}, \vec{\xi})$, defined in \eq{deviation}, of the spectral range of $\hat{H} - \hat{K}$ to the spectral range of $\hat{H}$ restricted to the target electron number subspace. (b) 1-norms of Pauli and DF LCUs of the original Hamiltonian, without BLISS. (c) Ratio of Pauli 1-norm of the Hamiltonian after BLISS to the Pauli 1-norm without BLISS. (d) Ratio of fermionic 1-norm of the Hamiltonian using a BLISS operator to the DF 1-norm without BLISS. The vertical gray lines denote the interval between $\Delta E / 2$ and $\Delta E^{(\text{ENS})} / 2$ for the original Hamiltonian $\hat{H}$. The horizontal gray ticks denote $\Delta E(\vec{\mu}, \vec{\xi}) / 2$ associated to FLR-BLISS. See Appendix \ref{appendix_glossary} for acronym definitions.}
 	\label{fig_1norms}
\end{figure*}

First, we assess the BLISS operators obtained from linear programming in terms of their ability to reduce the spectral range of the electronic Hamiltonian. To do this, we compare three quantities: (1) the spectral range $\Delta E$ of $\hat{H}$, (2) the spectral range $\Delta E(\vec{\mu}, \vec{\xi})$ of $\hat{H} - \hat{K}$, and (3) the spectral range $\Delta E^{(\text{ENS})}$ of $\hat{H}$ when restricted to the target electron number subspace. Since the spectral range is computationally difficult to calculate for systems of the size considered in this work, we approximate the largest and lowest energies using a modification of the Lanczos procedure, \cite{lanczos1950iteration} described in Appendix \ref{appendix_C}. We quantify the ability of a BLISS operator to reduce the spectral range with the rescaled difference between $\Delta E(\vec{\mu}, \vec{\xi})$ and $\Delta E^{(\text{ENS})}$ 
\begin{equation}
    D(\vec{\mu}, \vec{\xi}) = \frac{\Delta E(\vec{\mu}, \vec{\xi}) - \Delta E^{(\text{ENS})}}{\Delta E - \Delta E^{(\text{ENS})}}, \label{deviation}
\end{equation}
which is normalized such that $D(\vec{\mu}, \vec{\xi}) = 0$ when the spectral range is reduced to the theoretical lower bound: $\Delta E(\vec{\mu}, \vec{\xi}) = \Delta E^{(\text{ENS})}$, and $D(\vec{\mu}, \vec{\xi}) = 1$ when the spectral range is unmodified: $\Delta E(\vec{\mu}, \vec{\xi}) = \Delta E$. \Fig{fig_1norms}a  depicts the value of $D(\vec{\mu}, \vec{\xi})$ for the three methods to generate the BLISS operator described in Sec. \ref{theory_sec}. Although all three methods substantially reduce the spectral range for all systems considered, the FLR-BLISS method, for which the mean of $D(\vec{\mu}, \vec{\xi})$ is $0.04$, and the maximum of $D(\vec{\mu}, \vec{\xi})$ is $0.10$, most consistently achieves the largest reductions on average across the systems considered. The FFR-BLISS method, with a mean of $0.06$ and a maximum of $0.16$ for $D(\vec{\mu}, \vec{\xi})$, was not as successful at lowering the spectral range on average, but did achieve the lowest spectral range value for three of the systems considered. The LP-BLISS method, with a mean of $0.09$, and a maximum of $0.20$ for $D(\vec{\mu}, \vec{\xi})$, was the worse performing on average, but still achieved the lowest value for two of the systems considered.


\begin{figure*}
 	\centering
 	\includegraphics[width=0.75\textwidth]{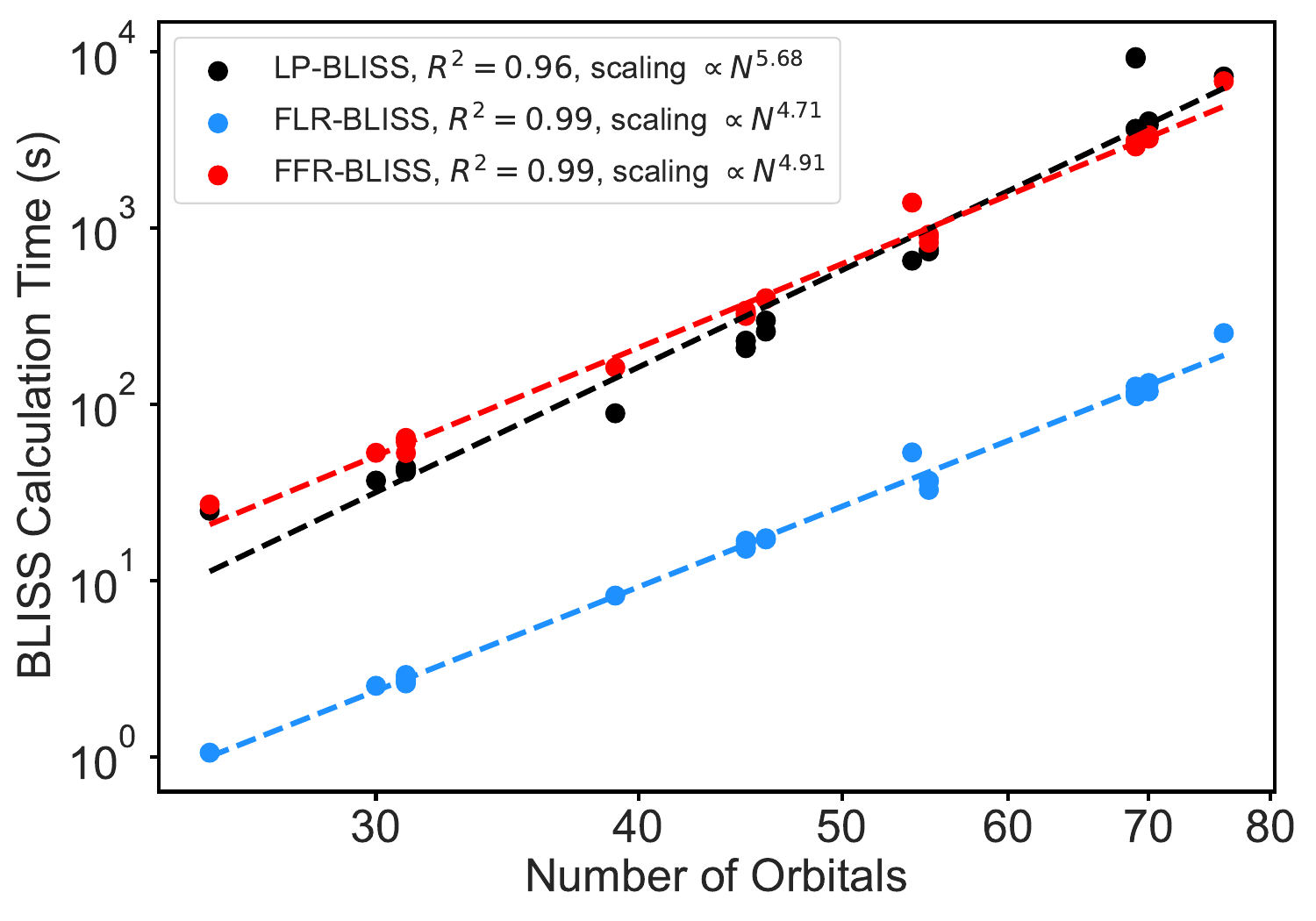}
 	\caption{Calculation time to generate the optimized BLISS parameters $\vec{\mu}, \vec{\xi}$ and modified one- and two- electron integrals of $\hat{H} - \hat{K}$ versus the number of orbitals. The dashed line is a fit to the data, with the coefficient of determination and exponent shown in the legend. See Appendix \ref{appendix_glossary} for acronym definitions.}
 	\label{fig_runtimes}
 \end{figure*}

We also assess the computational cost of generating the optimized BLISS operators and the transformed electronic Hamiltonian integrals. For this, we use the Niagara cluster hosted by SciNet. \cite{ponce2019deploying,loken2010scinet} SciNet is partnered with Compute Ontario and the Digital Research Alliance of Canada. Each node of Niagara  has 188 GiB of RAM and either 40 Intel ``Skylake'' 2.4 GHz cores or 40 Intel ``CascadeLake'' 2.5 GHz cores. All three methods to generate the BLISS operator scale polynomially in the system size, as they require, at most, polynomial-sized linear optimizations, in some cases after performing a double-factorization of the two-electron integrals. The run-times for generating the three types of BLISS operators considered in this work are shown in \fig{fig_runtimes}. LP-BLISS, which has an empirical scaling of $O(N^{5.68})$, had the worst scaling of all methods considered, and took 155 minutes to complete in the worst case. The FFR-BLISS method was marginally better, having an empirical scaling of $O(N^{4.91})$, and taking $113$ minutes in the worst case. However, we note that, unlike LP-BLISS, the FFR-BLISS operator is obtained by solving many small linear programs, as opposed to one large linear program. It can therefore be obtained via parallel programming, potentially allowing further speed up. Despite this, the cheapest method by far to generate the BLISS operator was FLR-BLISS, owing to the fact that the linear programs defining the FLR-BLISS operator all have analytical solutions. Although the empirical scaling of $O(N^{4.71})$ is comparable to the FFR-BLISS empirical scaling, we found that for all systems considered, generating and applying the FLR-BLISS operator took under 5 minutes. Therefore, of the three methods to obtain a BLISS operator considered in this work, FLR-BLISS is both the fastest and the most successful at reducing the spectral range. 


We now assess the effect of BLISS operators on the 1-norm of LCU decompositions of the electronic Hamiltonian. We analyze the results in terms of how much the 1-norm of a particular LCU is reduced as a consequence of the BLISS operator. The values of the Pauli 1-norm and DF 1-norm, without BLISS, for all systems considered, are shown in \fig{fig_1norms}b. On average, the DF LCU achieves a 3.9 fold reduction in the 1-norm compared to the Pauli LCU.

\Fig{fig_1norms}c depicts the ratio of the Pauli LCU 1-norm after applying BLISS to the Hamiltonian, to the Pauli LCU 1-norm of original Hamiltonian. LP-BLISS, which directly targets the Pauli LCU 1-norm, achieves a non-trivial reduction for all systems considered, with an average reduction of 23\%.  One notable system is FeMoco Lg, for which the reduction in Pauli LCU 1-norm achieved by LP-BLISS is 59\%, the largest seen in this work. Since LP-BLISS directly targets the Pauli LCU 1-norm, the LP-BLISS+Pauli LCU 1-norms represent the maximal reduction that can be achieved via optimizing the BLISS operator. Despite this, we still see a sizeable reduction in the 1-norm when pre-processing $\hat{H}$ using the FFR-BLISS and FLR-BLISS operators. The FFR-BLISS and FLR-BLISS values are similar for all systems considered, achieving 16\% and 17\% reductions on average respectively. However, as shown in the Supplementary Information, neither FFR-BLISS nor FLR-BLISS were able to outperform, on average, the usage of a simple symmetry shift operator for reducing the Pauli LCU 1-norm. Therefore, these results suggest that, in the regime where the classical cost of the Hamiltonian pre-processing is a bottleneck, one should use a symmetry-shift as a computationally inexpensive method to reduce the Pauli LCU 1-norm, rather than FFR-BLISS or FLR-BLISS. Note also that, in principle, one could simultaneously optimize the orbital basis in which $\hat{H}$ is expressed together with the BLISS operator to achieve a further reduction beyond what LP-BLISS can produce, \cite{koridonOrbitalTransformationsReduce2021,loaizaReducingMolecularElectronic2023a} but this requires a nonlinear optimization which incurs a substantial classical computational cost.

\Fig{fig_1norms}d depicts the 1-norms associated to fermionic LCUs based on the two approaches to incorporate the BLISS operator: (1) pre-processing $\hat{H}$ by applying a BLISS operator and subsequently performing double-factorization (LP-BLISS+DF, FLR-BLISS+DF), and (2) performing double-factorization and subsequently post-processing the fragments (DF+LRPS and DF+LRBS). We omit the results obtained from FFR-BLISS+DF, since this method was not competitive in 1-norms with the other pre-processing methods. Like with the Pauli LCU, we found that both the LP-BLISS+DF and FLR-BLISS+DF LCUs have lower 1-norms compared with DF LCU. We also see that DF combined with LP-BLISS gives a greater reduction in 1-norm (7.7 times, on average) than either LP-BLISS (1.3) or DF (3.9) alone, when compared with the Pauli 1-norm. For the post-processing methods, we found that, for all systems considered, DF+LRPS achieves a substantial reduction in the 1-norm compared with both DF and DF+LRBS. This suggests that constraining the BLISS operator to preserve the low-rank property is a better strategy for reducing the 1-norm than introducing additional optimization parameters which break the low-rank property. As shown in \fig{fig_1norms}d, DF+LRBS does not yield much improvement in 1-norms even compared to DF, and sometimes performs worse, which suggests that even the most generally optimized BLISS operator is not able to overcome the increase in 1-norm associated to breaking the low-rank property. When comparing the 1-norms of DF+LRPS and the pre-processing methods, the 1-norms are remarkably similar, with the DF+LRPS 1-norms coming within 8\% to the pre-processing 1-norms for all systems considered. Despite this similarity, DF+LRPS produces the lowest 1-norms (\fig{fig_1norms}d) for all but three of the systems considered in this study, and also produces the lowest average 1-norms of all methods. Also, as shown in the Supplementary Information, all methods, apart from DF+LRBS, outperformed, on average, the usage of a symmetry shift operator for reducing the DF 1-norm, which was not the case for the Pauli LCU 1-norms.

The two active spaces of FeMoco defined In Refs. \cite{reiherElucidatingReactionMechanisms2017} and \cite{liElectronicComplexityGroundstate2019} are commonly used systems to benchmark fault tolerant algorithms. Therefore, for these systems, we are able to compare our 1-norms to others reported in the literature, the results of which are shown in Table \ref{table_comparisons}. Compared with tensor hypercontraction (THC), \cite{leeEvenMoreEfficient2021} the DF+LRPS method produced a 45\% lower 1-norm for FeMoco Sm, and a 59\% lower 1-norm for FeMoco Lg. In the SCDF method, \cite{roccaReducingRuntimeFaultTolerant2024a} the LCU and the symmetry shift operator are optimized simultaneously, and indeed, the simultaneous optimization produces a 54\% lower 1-norm than the DF+LRPS method. In a more recent work, \cite{dekaSimultaneouslyOptimizingSymmetry2024} a method to combine the optimization of the DF decomposition together with the full BLISS operator was proposed, which we will refer to as the ``Simultaneously-optimized DF and BLISS'' (SDFB) method. SDFB currently achieves the best 1-norm for FeMoco Sm, beating DF+LRPS by 66\% However, both SCDF and SDFB methods require the usage of non-linear cost functions, and are thus much more computationally expensive than DF+LRPS, and currently there are no results for either method when applied to the larger FeMoco Lg system. Furthermore, given that DF+LRPS is a post-processing method, it can be applied on top of the SCDF or SDFB fragments, potentially yielding a further reduction in 1-norms.

\begin{table}
\begin{tabularx}{\columnwidth}{@{\extracolsep{\fill}} l c c c c c}
\toprule
          & DF     & DF+LRPS & THC    & SCDF & SDFB\\ \midrule
FeMoco Sm & 296.9  & 169.4   & 306.3  & 78.0 & 57.9    \\ \midrule
FeMoco Lg & 1174.0 & 487.5   & 1201.5 & $-$  & $-$     \\ \bottomrule
\end{tabularx}
\caption{Comparison of 1-norms of different fermionic LCU decompositions of the two FeMoco active space Hamiltonians. THC values were taken from Ref. \cite{leeEvenMoreEfficient2021}. SCDF values were taken from Ref. \cite{roccaReducingRuntimeFaultTolerant2024a}. SDFB values were taken from Ref. \cite{dekaSimultaneouslyOptimizingSymmetry2024}.}
\label{table_comparisons}
\end{table}

Combining BLISS techniques with DF produces 1-norms that are lower than the $\Delta E/2$ lower bound for many of the systems considered (see \fig{fig_1norms}d). It is thus clear that BLISS generally lowers the spectral-range of the Hamiltonian in such a way that DF can take advantage of the reduced range, which is not otherwise \textit{a priori} guaranteed. It also confirms that any technique to reduce the 1-norm beyond what is achieved in this work \textit{requires} a modification of the target Hamiltonian to reduce its spectral-range. Despite this, we find that the 1-norms achieved in this work are still quite distant from the lower-bound $\Delta E^{(\text{ENS})} / 2$ that is achievable in theory, even though $\Delta E(\vec{\mu}, \vec{\xi})/2$ obtained from FLR-BLISS is relatively close to $\Delta E^{(\text{ENS})}/2$, as shown in \fig{fig_1norms}d. As described in Sec. \ref{sec_fermionic_lcu}, the deviation of the DF 1-norm from $\Delta E/2$ is a consequence of the requirement to use multiple fragments to decompose the Hamiltonian. Therefore, to close the gap between the DF 1-norm after BLISS, and $\Delta E^{(\text{ENS})}/2$, one approach can be to obtain the low-rank fragments in a way that concentrates the highest weight terms on as few fragments as possible. Various approaches have been developed to achieve this \cite{yenCartanSubalgebraApproach2021,oumarouAcceleratingQuantumComputations2024,cohnQuantumFilterDiagonalization2021} via nonlinear optimization techniques. However, we note that the method used in this work to obtain the low-rank fragments, based on SVD of the two-body-integral tensor, already produce fragments with this property.

Although DF+LRPS and DF+LRBS are mutually exclusive with each other, they are not mutually exclusive with the various methods for generating BLISS operators that act directly on the Hamiltonian. That is, one can envision a method with three steps: (1) apply a BLISS operator to produce $\hat{H} - \hat{K}$, where $\hat{K}$ is obtained via any of the three BLISS techniques considered in this work; (2) obtain the LCU of $\hat{H} - \hat{K}$ via DF, and (3) post-process the resulting fragments using LRBS or LRPS. We found that, for most systems, such a combination does not reduce the 1-norms beyond what could be obtained from the methods presented in  \fig{fig_1norms}d, and, in the cases where there was an improvement, it was always less than 5\%. Overall, when comparing all fermionic methods, including ones which combine a pre- and post-processing method, DF+LRPS produces the largest average reduction in 1-norm compared with the DF LCU 1-norm. The reason combining methods does not improve the 1-norm in general is twofold. First, once an optimized BLISS operator is subtracted from the electronic Hamiltonian, the values of the optimized parameters for both DF+LRPS and DF+LRBS are much closer to $0$ than they otherwise would have been had a BLISS operator not been subtracted first. This implies that the fermionic methods are not able to reduce the 1-norm of the two-electron fragments much further once a BLISS operator is subtracted from the electronic Hamiltonian. Second, when applying DF+LRPS to the BLISS-modified electronic Hamiltonian, for most systems, the one-electron corrections (\eq{lrps_1e_correction}) increased the 1-norm of the one-electron Hamiltonian more than the sum of reductions in the 1-norms of the two-electron fragments associated with applying the LRPS. 

Lastly, the cost of block-encoding within the LCU framework is not only dependent on the 1-norm but also on the cost of implementing each unitary in the LCU decomposition of $\hat{H}$, as well as the total number of unitaries. The main contributor to the cost of unitaries is the total number of non-Clifford gates, for example, T gates, which are the most expensive gates to implement fault-tolerantly. \cite{gidneyEfficientMagicState2019} All BLISS procedures do not change appreciably the number or character of terms in $\hat H-\hat K$ compared to those of $\hat H$, and the number of unitaries in any LCU is almost the same for $\hat H-\hat K$ as for $\hat H$ and will depend on the method of LCU decomposition. \cite{loaizaMajoranaTensorDecomposition2024} Thus, all BLISS procedures do not change the number of T gates needed for one step of the LCU encoding, but by reducing the total number of needed steps through 1-norm reduction, they reduce the total number of T gates.

\section{\label{conclusions_sec}Conclusions}

In this work, we introduce several approaches for reducing the spectral range of the electronic Hamiltonian and the 1-norms of qubit and fermionic LCUs. All the developed methods use linear programming, which allows us to optimize the BLISS operator to achieve a global minimum in the LCU 1-norm for a given decomposition. Using linear programming also allows for a much faster determination of the BLISS operator compared to previous methods, which relied on nonlinear, gradient based optimization. The FLR-BLISS method was the most effective at reducing the spectral range of the Hamiltonian. On average, the spectral range of the modified Hamiltonians was reduced by 96\% toward the theoretical lower bound, corresponding to the spectral range of the original Hamiltonian in the number-conserving subspace of interest. It was also the fastest method to process the Hamiltonian with a BLISS operator of the methods considered in this work.


The main advantage when pre-processing the Hamiltonian with a BLISS operator is a highly efficient reduction in the spectral range, decreasing the lower bound of the 1-norm of LCU decompositions of the modified Hamiltonian compared with the original Hamiltonian. Thus, one can use BLISS processed Hamiltonians and achieve better results with various LCU decompositions that go beyond using linear combination of Pauli products. For Pauli LCU decompositions, LP-BLISS provides the global minimum that can be achieved in the 1-norm, for example, a 59\% reduction in Pauli LCU 1-norm for FeMoco with active space defined in Ref. \cite{liElectronicComplexityGroundstate2019}. The much computationally faster FLR-BLISS operator obtained in the fermionic algebra achieved, on average, 71\% of the reduction in the Pauli 1-norm compared with the LP-BLISS operator. We also calculated the 1-norms of fermionic LCUs based on a low-rank decomposition of the two-electron integral tensor obtained via DF. In this case, the DF + LRPS method produced the lowest 1-norms, while also being the fastest method for generating a fermionic LCU using BLISS techniques. Notably, DF + LRPS outperformed LP-BLISS+DF in generating lower 1-norms, as LP-BLISS is optimized for Pauli product LCUs, not for DF LCUs.

These points make a strong case for processing any Hamiltonian using BLISS techniques before doing block encoding. To make further improvements in the 1-norm cost there are two possible directions: 1) optimizing the LCU decomposition to drive the 1-norm closer to the spectral range of $\hat{H}-\hat{K}$ , 2) creating an effective Hamiltonian that has the same low-energy spectrum as the original electronic Hamiltonian but a lower spectral range. 

\section*{Acknowledgements}
The authors thank Nicole Bellonzi and Alexander Kunitsa for the nitrogen fixation catalyst Hamiltonians. S.P. and A.S.B. thank Ignacio Loaiza for helpful discussions.
J.T.C. thanks A.M. Rey for her hospitality during his visit to JILA at CU Boulder. 
S.P. and A.F.I. gratefully acknowledge financial support from NSERC and DARPA.
This research was partly enabled by Compute Ontario (computeontario.ca) and the Digital Research Alliance of Canada (alliancecan.ca) support. Part of the computations were performed on the Niagara supercomputer at the SciNet HPC Consortium. SciNet is funded by Innovation, Science, and Economic Development Canada, the Digital Research Alliance of Canada, the Ontario Research Fund: Research Excellence, and the University of Toronto.

\appendix
\section{Glossary}
\label{appendix_glossary}

\noindent \textbf{DF} Double Factorization.

\noindent \textbf{LCU}  Linear Combination of Unitaries.

\noindent \textbf{BLISS}  Block Invariant Symmetry Shift, Ref. \cite{loaizaBlockInvariantSymmetryShift2023}.

\noindent \textbf{LP-BLISS}  Linear-Programming BLISS, Section \ref{theory_pauli_subsec}.

\noindent \textbf{FLR-BLISS}  Fermionic-Low-Rank BLISS, Section \ref{section_individual_ferm_frag}.

\noindent \textbf{FFR-BLISS}  Fermionic-Full-Rank BLISS, Section \ref{section_individual_ferm_frag}.

\noindent \textbf{DF+LRPS}  Double-Factorization with Low-Rank Preserving Shifts, Section \ref{section_individual_ferm_frag}.

\noindent \textbf{DF+LRBS}  Double-Factorization with Low-Rank Breaking Shifts, Section \ref{section_individual_ferm_frag}.

\noindent \textbf{[Method] + Pauli} Pauli decomposition after [Method] applied.

\noindent \textbf{[Method] + DF} DF decomposition after [Method] applied.

\noindent \textbf{SCDF} Symmetry-Compressed Double Factorization, Ref. \cite{roccaReducingRuntimeFaultTolerant2024a}.

\noindent \textbf{THC} Tensor hypercontraction, Ref. \cite{leeEvenMoreEfficient2021}.

\noindent \textbf{SDFB} Simultaneously-optimized DF and BLISS, Ref. \cite{dekaSimultaneouslyOptimizingSymmetry2024}.

\section{\label{appendix_A}1-norm minimization as a linear program}

All optimizations of the 1-norm considered in this work can be phrased as a minimization problem in which one obtains $\min_{\vec{x} \in \mathbb{R}^n} |A\vec{x} - \vec{b}|_1$ for some given $A \in \mathbb{R}^{n \times m}$ and $\vec{b} \in \mathbb{R}^{m}$. This can be written as a linear program by introducing a new variable $\vec{y} \in \mathbb{R}^m$ and writing the minimization problem as 
\begin{equation}
    \min_{(\vec{x},\vec{y}) \in \mathbb{R}^{n+m}} \sum_{i=1}^{m} y_i
\end{equation}
subject to the constraint $\vec{y} \geq |A\vec{x} - \vec{b}|_1$. This constraint can be written in terms of two linear constraints: 
$\vec{y} \geq \pm (A\vec{x} - \vec{b})$, or, in the matrix form as
\begin{equation}
    \begin{bmatrix} A & -I \\ -A & -I\end{bmatrix}\begin{bmatrix} \vec{x} \\ \vec{y} \end{bmatrix} \leq \begin{bmatrix} \vec{b} \\ -\vec{b} \end{bmatrix},
\end{equation}
which is the linear program form.

\section{\label{appendix_B}Details of low-rank preserving coefficient shift}

In this section, we show that using the DF+LRPS method allows one to obtain a decomposition of $\hat{H}_{2e}$, up to a BLISS operator, a one-electron operator, and a constant. For simplicity of notation, we define the following:

\begin{align}
    \hat{\ell}(\vec{\epsilon}) &= \sum_{i\sigma} \epsilon_i \hat{n}_{i\sigma}\\
    \hat{b}_{N_e}(\vec{\epsilon}) &= \sum_{ij\sigma\tau} \left(\frac{\epsilon_i + \epsilon_j}{2}\right) \hat{n}_{i\sigma}\hat{n}_{j\tau} - N_e\ell(\vec{\epsilon}).
\end{align}
$\hat{\ell}(\vec{\epsilon})$ corresponds to a generic spin-symmetric linear combination of number operators, whose square $\hat{\ell}^2(\vec{\epsilon})$ is present in the low-rank fragments, and $\hat{b}_{N_e}(\vec{\epsilon})$ corresponds to the number operator polynomial present in the BLISS operator shown in \eq{fermionic_bliss}. Then, after some straightforward algebra, one can relate the number operator polynomial present in a given DF+LRPS fragment to the number operator polynomial $\hat{\ell}^2(\vec{\epsilon})$ present in the associated DF fragment as follows:
\begin{align}
    \left[\sum_{i\sigma} (\epsilon_i - \phi) \hat{n}_{i\sigma}\right]^2 &= \hat{\ell}^2(\vec{\epsilon}) - 2\phi N_e \hat{\ell}(\vec{\epsilon}) + \phi^2 N_e^2\nonumber\\
    &+ \big(\phi^2 (\hat{N}_e^2 - N_e^2) - 2\phi \hat{b}_{N_e}(\vec{\epsilon})\big).
\end{align}
Therefore, given a low-rank fragment of the electronic Hamiltonian in DF
\begin{align}
    \hat{H}^{(\alpha)} &= \hat{U}^{(\alpha)\dagger} \left[\sum_{i\sigma} \epsilon_i^{(\alpha)} \hat{n}_{i\sigma}\right]^2 \hat{U}^{(\alpha)}\nonumber\\
    &=\hat{U}^{(\alpha)\dagger}\hat{\ell}^2(\vec{\epsilon}^{\hspace{0.2em}(\alpha)}) \hat{U}^{(\alpha)},
\end{align}
we can write the modified fragment $\hat{H}^{(\alpha)}(\phi^{(\alpha)})$ obtained in the DF+LRPS method as follows: 
\begin{align}
    \hat{H}^{(\alpha)}(\phi^{(\alpha)}) &= \hat{U}^{(\alpha)\dagger}\left[\sum_{i\sigma} (\epsilon_i^{(\alpha)} - \phi^{(\alpha)}) \hat{n}_{i\sigma}\right]^2 \hat{U}^{(\alpha)}\\
    &= \hat{H}^{(\alpha)} - \hat{S}_{1e}^{(\alpha)} + \phi^{(\alpha)^2} N_e^2 \nonumber\\
    &+ \hat{K}^{(\alpha)}\left(\phi^{(\alpha)^2}, -2\phi^{(\alpha)}\vec{\epsilon}^{\hspace{0.2em}(\alpha)}\right),
\end{align}
where $\hat{K}^{(\alpha)}\left(\phi^{(\alpha)^2}, -2\phi^{(\alpha)}\vec{\epsilon}^{\hspace{0.2em}(\alpha)}\right)$ is the BLISS operator as defined in \eq{fermionic_bliss}, and $\hat{S}_{1e}^{(\alpha)}$ is a one-electron correction:
\begin{equation}
    \hat{S}_{1e}^{(\alpha)} = 2\phi^{(\alpha)} N_e \hat{U}^{(\alpha)\dagger}\hat{\ell}(\vec{\epsilon}^{\hspace{0.2em}(\alpha)}) \hat{U}^{(\alpha)}. \label{lrps_1e_correction}
\end{equation}

\section{\label{appendix_C} Estimating the spectral range with a truncated Lanczos method}

To estimate Hamiltonian spectral ranges within the entire Fock-space and those that are restricted to a particular electron-number-subspace we used the 
Lanczos algorithm \cite{lanczos1950iteration} with an additional truncation of vectors in the Krylov space. We work within the diagonalized one-electron operator $\hat{H}_{1e}$ orbital frame. This allows us to generate the initial state as a single Slater determinant $\ket{\psi_{\text{init}}}$ with $N_e$ occupied spin-orbitals corresponding to lowest (highest) orbital energies for obtaining the lowest (highest) eigen-states. To obtain the Fock-space spectral range, we repeat this procedure to all electron-number-subspaces.  The difficulty with generating the Krylov space vectors is growing size of vectors after each application of $\hat{H}_e$. To reduce the computational cost, we simplify the state in the $k^{\text{th}}$ iteration by retaining only $5k$ Slater determinants with largest coefficients in absolute value. After this truncation, the state is orthogonalized with the previous states using the Gram-Schmidt procedure. The algorithm is terminated when the residual vector after the Gram-Schmidt procedure has a 2-norm lower than $10^{-5}$; otherwise the residual vector is normalized and the next iteration starts. At the end of the procedure, the Hamiltonian is diagonalized within the subspace spanned by Krylov space vectors to identify the extreme eigenvalue states. In spite of the truncation the approach is variational and thus gives a lower bound on the actual spectral range. 

\bibliography{library}


\end{document}